# Impact of gigahertz and terahertz transport regimes on spin propagation and conversion in the antiferromagnet IrMn

O. Gueckstock[1,2,**], R. L. Seeger[3,**], T. S. Seifert[1,2], S. Auffret[3], S. Gambarelli[4], J. N. Kirchhof[1], K. I. Bolotin[1], V. Baltz[3*], T. Kampfrath[1,2], L. Nádvorník[5,*]

1. Department of Physics, Freie Universität Berlin, 14195 Berlin, Germany
2. Department of Physical Chemistry, Fritz Haber Institute of the Max Planck Society, 14195 Berlin, Germany
3. Univ. Grenoble Alpes, CNRS, CEA, Grenoble INP, IRIG-Spintec, F-38000 Grenoble, France
4. Univ. Grenoble Alpes, CNRS, CEA, SYMMES, F-38000 Grenoble, France
5. Faculty of Mathematics and Physics, Charles University, 121 16 Prague, Czech Republic

[*] Authors to whom correspondence should be addressed: nadvornik@karlov.mff.cuni.cz, vincent.baltz@cea.fr

[**] contributed equally to this work

## Abstract

Control over spin transport in antiferromagnetic systems is essential for future spintronic applications with operational speeds extending to ultrafast time scales. Here, we study the transition from the gigahertz (GHz) to terahertz (THz) regime of spin transport and spin-to-charge current conversion (S2C) in the prototypical antiferromagnet IrMn by employing spin pumping and THz spectroscopy techniques. We reveal a factor of 4 shorter characteristic propagation lengths of the spin current at THz frequencies (~ 0.5 nm) as compared to the GHz experiments (~ 2 nm). This observation may be attributed to different transport regimes. The conclusion is supported by an extraction of sub-picosecond temporal dynamics of the THz spin current. We identify no relevant impact of the magnetic order parameter on the S2C signals and no scalable magnonic transport in the THz experiments. A significant role of the S2C originating from the interfaces between the IrMn and magnetic or non-magnetic metals is observed which is much more pronounced in the THz regime and opens the door for optimization of the spin control at ultrafast time scales.

Antiferromagnetic spintronic devices[1] provide many advantages such as robustness against external magnetic fields, a higher memory bit integration, two orders of magnitude faster manipulation of the magnetic order and new topological phenomena[2,3]. Their functionalities include pseudospin dynamics of magnons[4] and a wide spectrum of applications like memory[5-9], spin logic[10] and terahertz (THz) emission devices using pinning of a hard magnetic layer[11] or gradual reorientation of the Néel-vector[12]. To exploit these advantages, we need to control (i) the injection, (ii) transport and (iii) conversion of the spin angular momentum in antiferromagnetic materials.

A model metallic antiferromagnet (AF) is IrMn in which spin-transfer effects[13], spin-orbit effects[14] and ferromagnetic reversal by spin Hall torques[15,16] have been exploited. It was shown that the AF ordering plays no significant role for spin transport in IrMn polycrystalline films[1,14]. This behavior was suggested to arise from the different direction of the moments averages out any anisotropic spin-relaxation contribution due to the magnetic order. Interestingly, the fact that the spin transport does not depend on the magnetic order parameter means that they can be obtained from the paramagnetic state and applied to the technologically relevant AF case, in line with earlier strategies used for AF spintronics[1,14]. In addition, regarding (i), an enhancement of the spin injection in IrMn by spin pumping due to spin fluctuations around the Néel temperature ($T_\mathrm{N}$) at GHz frequencies may be possible[17]. The origin of the effect lies in the direct link between the spin mixing conductance and the linear dynamic spin susceptibility[18,19].

In terms of (ii), in IrMn and structurally similar FeMn, two types of spin transport – electronic and magnonic – may exist at GHz frequencies[20,21] as indicated by spin-pumping techniques. Experiments in FeMn[20] suggest different spin-current penetration depths in both regimes (1 and 9 nm, respectively). Lastly, regarding (iii), spin-to-charge-current conversion (S2C) in IrMn was studied at DC and AC frequencies, giving a spin Hall angle of a few percent[14,22-24]. A non-monotonic contribution to the temperature-dependent S2C signal due to nonlinear spin susceptibilities around $T_\mathrm{N}$ may also be possible, although not demonstrated so far, similar to findings in the PdNi weak ferromagnet[25]. This contribution relates to a different term that is the second order nonlinear dynamic susceptibility. To utilize the full potential of antiferromagnetic spin transport, spin currents have to be transferred to the ultrafast regime that matches the dynamics of the antiferromagnetic order parameter. So far, only a few recent studies focused on the spin transport at terahertz (THz) frequencies[26-29].

In this paper, we explore the ultrafast (THz) spin injection, transport and S2C in $Ir_{20}Mn_{80}$ and directly compare them with transport experiments in the GHz range in equivalent samples. First, our results indicate a change in the nature of the spin transport when transiting from the GHz to the THz regime. Second, we show that S2C in IrMn at THz frequencies reaches similar efficiencies as in the GHz range. Interestingly, our observation suggests a strong influence of the interfaces between IrMn and the heavy-metal or the metallic magnet on the resulting in-plane charge current that is significantly more pronounced in the THz regime.

Our methodology is based on measuring S2C of spin currents injected from a layer of ferromagnetic $Ni_{81}Fe_{19}$ (F) into a bilayer of $Ir_{20}Mn_{80}$ (AF) and non-magnetic metal (N) [see Fig. 1(a,b)]. Spin angular momentum is injected in two different frequency ranges by (i) ferromagnetic spin pumping at 9.6 GHz (defined by the ferromagnetic resonance of NiFe), using a continuous-wave electron paramagnetic resonance spectrometer fitted with a three-loop-two-gap resonator[30] [Fig.1(a)], and (ii) ultrafast spin-voltage generation in NiFe at 0.1-30 THz[31-33] by an optical femtosecond pump pulse [Fig.1(b)]. In both techniques, the resulting out-of-plane spin current density $j_\mathrm{s}(\omega,z)$ is converted to an in-plane charge current $I_\mathrm{c}(\omega)$ by the local (layer-dependent) spin Hall angle $\theta(\omega,z)$, thus generating a detectable electric field $E$. In the frequency domain, the complex-valued field amplitude is given by

$$E(\omega) = Z(\omega)I_\mathrm{c}(\omega) = eZ(\omega)\int \mathrm{d}z\, j_\mathrm{s}(\omega,z)\,\theta(\omega,z) \qquad (1)$$

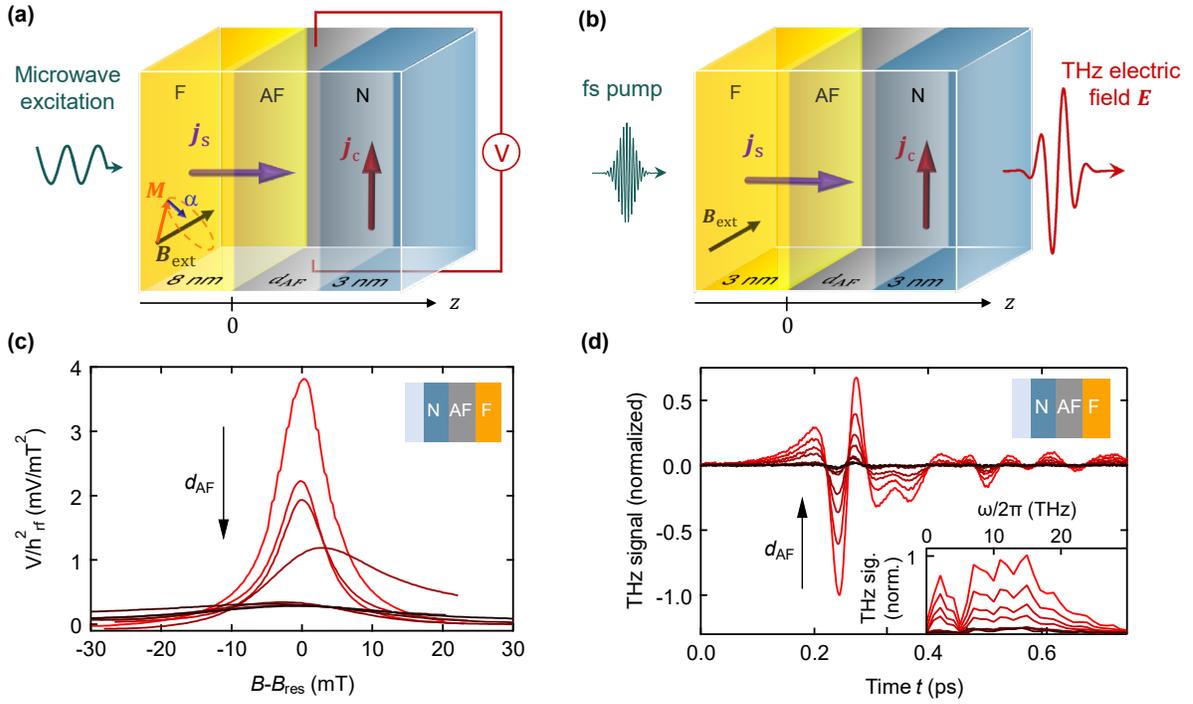

**FIG. 1. | Measuring inverse spin Hall effect at GHz and THz frequencies. (a)** Schematic of the GHz experiment. A microwave magnetic field (amplitude $h_{rf}$ ~ 0.05 mT, frequency 9.6 GHz) triggers the precession of magnetization in a magnetic layer (F = NiFe, thickness of 8 nm) and, due to spin pumping, launches a periodic spin current $j_s$ through the antiferromagnetic layer IrMn (AF, thickness $d_{AF}$) into a heavy metal layer (N, 3 nm) where it is converted into a detectable DC charge current $j_c$ via the inverse spin Hall effect. We note that the generated electric field is constant over the entire thickness of the thin-film stack. **(b)** The analogous experiment performed at THz frequencies. A femtosecond optical pulse triggers an ultrafast $j_s$ between the magnetic (F, 3 nm) and the AF layer. The converted $\boldsymbol{j_c}$ serves as a source of an emitted THz pulse. **(c, d)** Typical raw experimental data, illustrated here by N = Pt: normalized voltage $V/h_{rf}^2$ in the GHz (c) and the electro-optical signal in the THz (d) experiments for different $d_{AF}$ (black arrows indicate increase of $d_{AF}$). All waveforms in (d) were normalized by the amplitude corresponding to $d_{AF} = 0$. Inset: amplitude spectrum of the corresponding THz temporal waveforms.

Here, $z$ is the coordinate along the sample normal [see Fig. 1(a), (b)], $\omega$ is the angular frequency and $I_c(\omega)$ denotes the sheet charge current. $E(\omega)$, related to $I_c$ through the total sample impedance $Z(\omega)$, is detected (i) directly by electrical contacts on the sample, and (ii) contact-free by electro-optic sampling[34, 35] of the emitted THz pulse with a co-propagating probe pulse (0.6 nJ, 10 fs) in a 10μm-thick ZnTe(110) crystal under ambient conditions. We note that $Z$ is $z$-independent and represents the total impedance (i.e., the inverse of the sum of conductances of all layers). The electric field is constant and equals $E(\omega)$ across the thin-film stack because it propagates through the stack several times due to back reflections on sample boundaries[36] (see Supplementary Fig. 6 for more details).

To investigate the propagation of $j_s(z)$ in both frequency regimes, we study thickness-dependent series of samples in the form of trilayers N|AF|F and F|AF|N. Each of them consists of a F = NiFe with thicknesses of 3 nm and 8 nm for THz and GHz experiments, respectively. The AF layer is $Ir_{20}Mn_{80}$ with varying thickness $d_{AF}$ ranging from 0 nm up to 12 nm with a paramagnetic-to-antiferromagnetic phase transition expected at $d_{AF} \approx 2.7$ nm at room temperature[17]. Finally, the sample structures contain a heavy metal layer with N = Pt, W or Ta (all 3 nm). All samples are deposited on thermally oxidized Si on glass substrates with thicknesses of Si(0.3mm)|SiO$_2$(500nm) and SiO$_2$(0.5mm) for GHz and THz experiments, respectively. A 2-

nm-thick Al cap was deposited on all samples to form a protective AlOx film after oxidation in air. We note that the different thicknesses of the F layer serve to increase the impedance and, thus, increase the emitted THz amplitudes [Eq. (1)] or to reduce damping and subsequently increase spin injection efficiency in the spin pumping experiments[37, 38]. The impact of the F-dependent spin injection efficiency on the detected signals is removed by a normalization procedure described below.

Typical raw signals from the GHz and THz experiment are shown in Fig. 1(c)-(d) for various values of $d_{AF}$. In both experiments, the signal amplitudes decrease with increasing $d_{AF}$. The bandwidth of the THz setup is large enough to resolve sub-picosecond dynamics of the THz emission signal [Fig. 1(d) inset]. We note that the additional oscillations after the main pulse in the THz raw data [Fig. 1(d)] arise from water vapor absorption[39]. We also note that the GHz raw data gradually evolves from a Gaussian- to a Fano-like shape as $d_{AF}$ increases because for thick AF layers the S2C predominantly takes place inside the AF layer, which is relatively weak than the initial large S2C in the N layer, as detailed below (see also Supplementary Figs. S1 and S2).

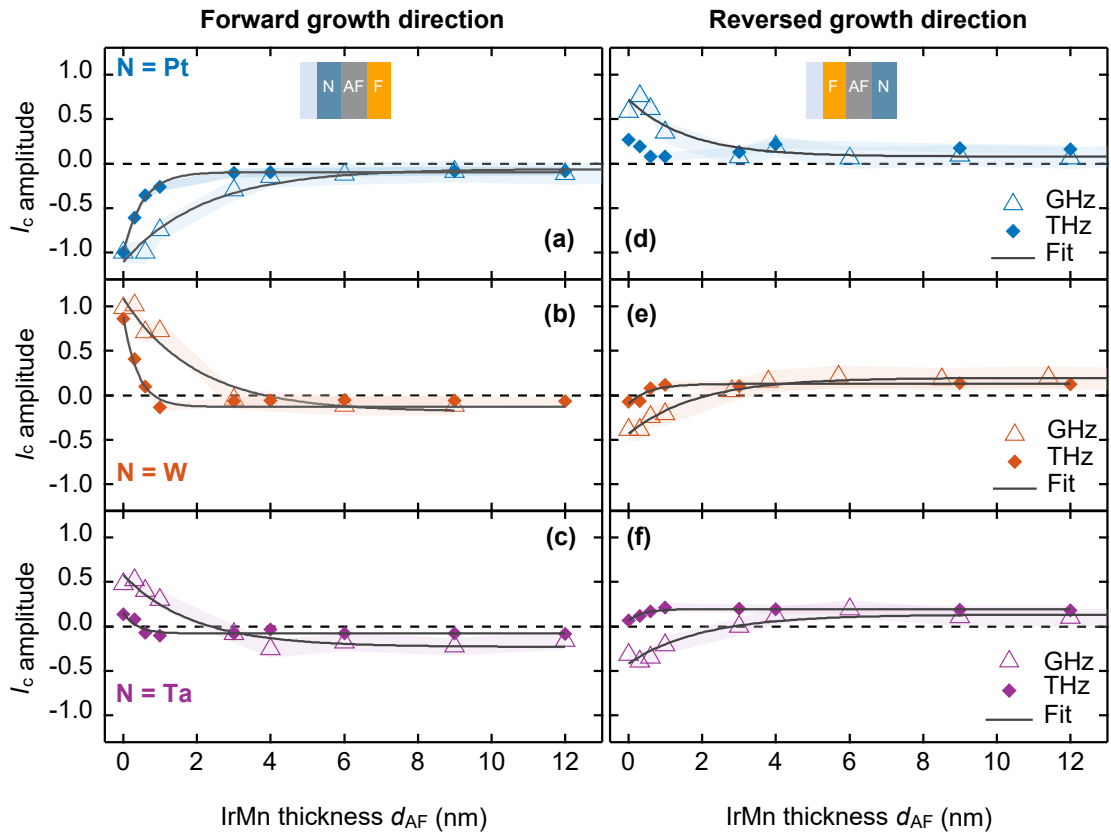

**FIG. 2. | Impact of IrMn thickness on GHz and THz charge current. (a)** Amplitude of charge currents $I_c$ as a function of the IrMn thickness in N|IrMn|NiFe for N = Pt, **(b)** N = W and **(c)** N = Ta layer at both frequency ranges (GHz: open triangles, THz: closed diamonds). The data are normalized to $I_c$ amplitudes obtained from Pt|IrMn($d_{AF} = 0$)|NiFe in GHz and THz sets, resulting in the effective spin-to-charge current conversion (S2C) efficiency relative to Pt|NiFe. **(d–f)** Same as (a–c), but for reversely grown NiFe|IrMn|N stacks. The data is normalized to account for thickness-dependent photonic and electronic effects unrelated to S2C. Errors are comparable to symbol sizes. The thick solid lines are guides to the eye. Fits (dark-gray thin solid lines) are offset mono-exponentials, giving the characteristic propagation lengths in IrMn. The fit values are summarized in Tab. 1 and Tab. S1.

To remove trivial $d_{AF}$-dependent photonic and electronic effects and, thus, make the data from the various samples directly comparable, we normalize[40] the signals by the independently measured $Z(\omega, d_{AF})$ and the absorbed powers of optical laser pulses or microwave GHz excitation (values for all samples are summarized in Supplementary Tab. S2), and take the root-mean square of the signals. The output of this procedure is the sheet charge-current amplitude $I_c$ normalized by the excitation power in the respective frequency range. We also remove all method-specific impacts on the measured signal (e.g., the effect of different thickness of F layer) by normalizing the GHz and THz data sets to Pt|IrMn(0 nm)|NiFe at the respective frequency range. The raw GHz voltage [Fig. 1(c)] was further treated to obtain its symmetric component as described in Fig. S1 and Ref. [30]. The resulting signals, shown in Fig. 2, directly capture the $d_{AF}$-dependence of $I_c$, which is a measure of the spin current $I_S$ and the S2C efficiency [Eq. (1)]. The underlying raw data sets are provided in Fig. S2.

We first analyze qualitatively the data in the "forward-grown" samples N|IrMn|NiFe [Fig. 2 (a)-(c)]. In both the GHz and THz regime, we observe a change in the signal polarity at $d_{AF} = 0$ when varying the N layer material, consistent with the sign and approximately the amplitudes of $\theta_N$ known from literature ($\theta_{Pt} > 0$ and $\theta_W, \theta_{Ta} < 0$)[41]. With increasing $d_{AF}$ and for a fixed N layer material, the signal decreases and, in the thick limit ($d_{AF} > 5$ nm), saturates at approximately the same value for all THz and GHz experiments. The thick-limit values are also consistent with THz and GHz signals from control bilayer samples of NiFe|IrMn with $d_{AF} = 12$ nm (see supplementary Fig. S5). The striking observation is the different rate of signal decay in both regimes which can be, in general, understood as a consequence of the finite propagation length of $j_s(z)$.

The accurate modeling of $j_s(z)$ in multilayers is typically a complicated task and requires the determination of many unknown parameters such as the spin mixing conductance of each interface[42]. To compare GHz and THz regime, we simplify the model by neglecting the back-reflections of $j_s(z)$ and consider the IrMn layer a simple exponential spin-current attenuator[32, 43], as illustrated for bilayer and trilayers by the sketches in Fig. 3. Consequently, the total sheet charge-current $I_c$ from Eq. (1) can be separated into contributions of three individual layers and two interfaces:

$$I_c(d_{AF}) = I_{c,AF} + I_{c,N} + I_{c,a} \approx I_{s,0}\left[(\lambda\theta^*)_{AF} + (\lambda\theta^*)_N e^{-d_{AF}/\lambda_{AF}}\right] + I_{c,a}. \quad (2)$$

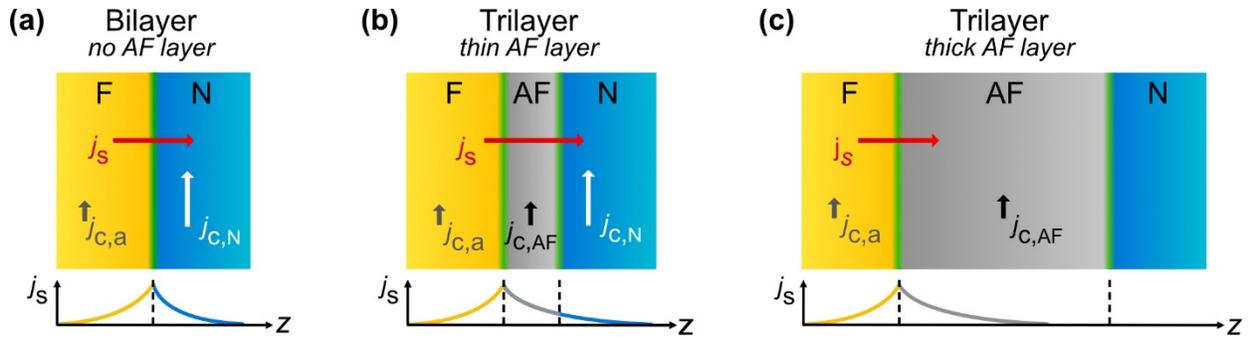

**FIG. 3 | Spin-to-charge current conversion scenarios.** The spin current density $j_s$ is generated inside the F layer and propagates to the AF|N substack. In general, S2C, leading to a sheet charge current $j_{c,i}$, can arise in all metallic layers $i$. **(a)** Without AF layer: S2C in N, i.e. $j_{c,N}$, dominates over $j_{c,a}$, which includes S2C in the F layer (yellow) or the interface(s) (thin green layers). **(b)** For thin intermediate layers of AF ($d_{AF} < \lambda$), an additional term $j_{c,AF}$ appears due to S2C inside the AF layer. The N layer still serves as an efficient spin-to-charge converter. **(c)** For thick intermediate layers of AF ($d_{AF} > \lambda$), $j_s$ does not reach the N layer and the S2C process is dominated by $j_{c,AF}$ or $j_{c,a}$. The sketches below the sample schematics illustrate the propagation length $\lambda$ of $j_s$ as a function of the sample thickness $z$ to help identify layers that can, in principle, contribute to the overall S2C process **[R1.C1a]**

|  | N | $\lambda_{\text{IrMn}}$ (nm) at 0.5-30 THz | $\lambda_{\text{IrMn}}$ (nm) at 9.6 GHz |
|---|---|---|---|
| **N \| IrMn \| NiFe** (forward-grown) | Pt | 0.6 ± 0.1 | 2.2 ± 0.5 |
|  | W | 0.4 ± 0.1 | 2.0 ± 0.7 |
|  | Ta | 0.4 ± 0.1 | 1.9 ± 0.5 |
| **NiFe \| IrMn \| N** (reversed-grown) | Pt | – | 1.6 ± 0.7 |
|  | W | 0.5 ± 0.2 | 1.9 ± 0.3 |
|  | Ta | 0.4 ± 0.1 | 2.0 ± 0.7 |

**TABLE 1. | Spin current characteristic lengths $\lambda$ as fitted from data shown in Fig. 2.** Fits are offset mono-exponential functions specified in the main text. The errors are obtained from fitting statistics and repeated experiments. The values of the other fit parameters $y_0$ and $y_1$ are summarized in the supplementary Tab. S1.

Here, $I_{s,0}$ is the total initial spin current launched by the excitation, $\lambda_i$ is the characteristic propagation length for the spin current in the corresponding layer $i = \text{F, AF, N}$, and $\theta_i^*$ the effective spin Hall angle which includes all possible effects of spin memory loss (not shown in Fig. 3 for simplicity) and spin mixing conductance between the layers[44]. The last term $I_{c,a} = I_{c,F} + I_{c,I}$ stands for an additional sheet charge current originating from S2C in the ferromagnetic layer and both interfaces[40]. Note that due to the simplifications, $\lambda_i$ cannot be rigorously taken as the spin diffusion length but rather serves as a quantity to compare spin transport in both frequency regimes. Similarly, we can view the quantity $(\lambda\theta^*)_i$ as the efficiency of the S2C that characterizes the practically achievable conversion in the layer including all mentioned spin injection losses.

We see that the model explains well the data in Fig. 2(a)-(f) (compare to Fig. 3). In the thin limit ($d_{AF} \lesssim \lambda_{AF}$), the bulk S2C $(\lambda\theta^*)_N$ is expected to dominate the S2C of the whole stack[14, 22-24] and the signal exponentially decreases with $d_{AF}$. For the thick limit ($d_{AF} > \lambda_{AF}$), the contribution from the N layer becomes negligible and the other, relatively small terms $I_{c,AF}$ and $I_{c,a}$ [Eq. (2)] start dominating in the signal. Within this limit, the negligible role of the N layer is verified by the mentioned N-free control bilayers NiFe|IrMn($d_{AF} = 12$ nm) (Fig. S5).

On a quantitative level, we use the model and fit the data in Fig. 2 by a offset mono-exponential function[32, 43] $y(d_{AF}) = y_1 e^{-d_{AF}/\lambda_{AF}} + y_0$ where $y_1$ and $y_0$ stand for the relative conversion $(\lambda\theta^*)_N/(\lambda\theta^*)_{Pt}$ and the sum of all remaining relative S2C, respectively [Eq. (2)]. The obtained values are summarized in Tab. 1 and Tab. S1. We remind that the data shown in Fig. 2 are normalized to the signal from the Pt|NiFe reference sample in the respective frequency range. The average relative efficiency of the S2C in the thick limit $y_{0,\text{THz}} \approx (8 \pm 1)\%$ and $y_{0,\text{GHz}} \approx (10 \pm 3)\%$ in the THz and GHz regimes, respectively, are reaching consistently similar values. We can interpret the findings as a demonstration that the spin-current injection and propagation in IrMn are operative at ultrafast time-scales. In the thin AF limit, the layer behaves like a mono-exponential spin current attenuator in the ultrafast THz regime[32, 43], qualitatively same as in the established GHz experiments[1, 45]. In the thick AF limit, when the ultrafast spin current does not reach the N layer, the role of IrMn as an attenuator changes to a converter, and the THz S2C efficiency signals saturate at very close averaged values as in the reference GHz measurements.

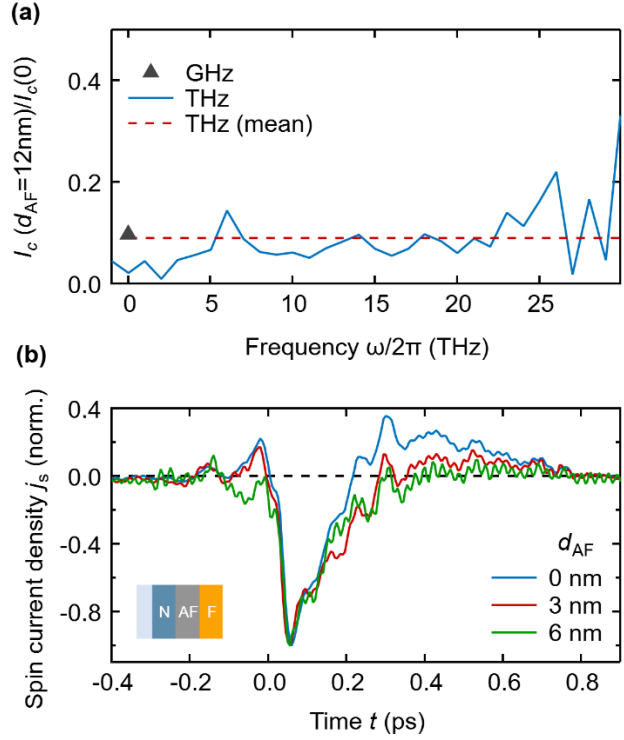

**FIG. 4. | Broadband THz charge and spin currents. (a)** Frequency dependence of the charge currents $I_c(d_{AF} = 12\,\text{nm}, \omega)/I_c(0, \omega)$ for the thick limit normalized to the Pt|NiFe reference sample ($d_{AF} = 0$). The data is extracted from the Pt|IrMn|NiFe series shown in panel Fig. 2 (a). The dotted red line depicts the mean value from the THz experiment corresponding to S2C in the thick limit (fitting parameter $y_{0,\text{THz}}$, see Tab. S1) and the triangle symbol is the same quantity in the GHz range. **(b)** Comparison of the extracted THz spin current $j_s(t)$ for $d_{AF} = 0$ (blue curve), 3 nm (red curve) and 6 nm (green curve), all normalized to peak value $-1$ for better comparability.

We note that a possible contribution to the THz emission signal originating from magnetic dipole radiation[33] it is usually an order of magnitude smaller and we, correspondingly, do not observe any significant THz signal contributions that are even upon sample reversal. Another source of the S2C signal may be the conversion in the F layer $I_{c,F}$. Although this contribution is typically neglected in GHz experiments[1], we should take the value of $y_0 \approx 8-10\%$ only as an upper bound of $(\lambda \theta^*)_{\text{IrMn}}/(\lambda \theta^*)_{\text{Pt}}$, i.e., the practically achievable total conversion signal in heterostructures including IrMn compared to Pt.

In addition to time-averaged values, the high temporal resolution of the THz experiment allows us to extract the $\omega$ dependence of $y_0$ taken from the THz data in Fig. 2(a). We observe a flat response between 0.5 and 30 THz and approximately the same value as in the GHz range [Fig. 4(a)]. The good agreement of GHz and THz S2C efficiency in both thickness limits is consistent with previous studies that compared THz and low-frequency regimes of spin-orbit-coupling-based effects[32, 46, 47].

As $y_0 \ll y_1$ in most cases, the IrMn layer behaves like a simple spin-current attenuator, and we can justify the mono-exponential approximation in Eq. (2). Using the fitting values from Tab. 1, we obtain the mean $\lambda_{\text{AF,THz}} = 0.5 \pm 0.1$ nm and $\lambda_{\text{AF,GHz}} = 1.9 \pm 0.6$ nm, averaged over stacks with different N. Except for the THz data in Fig. 2(d), we also do not observe any significant irregularities in the GHz and THz signals around the ordering thickness at $d_{AF} \approx 2.7$ nm at room temperature. We note that the pump laser pulse used in the THz experiments typically heats the electrons transiently by about 50-100 K[19], which would imply only a slight increase of the ordering thickness to $d_{AF} \approx 3.2-3.6$ nm[1, 17].

Interestingly, the factor of 4 between $\lambda_{\mathrm{AF,THz}}$ and $\lambda_{\mathrm{AF,GHz}}$ may indicate a different regime of spin transport in the THz and GHz range. To test this hypothesis, we take advantage of the time-resolved nature of the THz experiment and extract the ultrafast spin-current dynamics of $j_s(t)$ from the THz signals from Pt|NiFe [blue curve in Fig. 4(b)] and Pt|IrMn($d_{\mathrm{AF}}$)|NiFe with $d_{\mathrm{AF}} = 3$ and 6 nm (red and green curve), i.e., at $d_{\mathrm{AF}}$ where the IrMn is already antiferromagnetically ordered and allows for electronic and magnonic spin current. Data for more $d_{\mathrm{AF}}$ are displayed in Fig. S3. In all samples, the extracted $j_s(t)$ peaks at the same time and follows very similar dynamics as reported in previous works on fully metallic bilayer stacks (like F|Pt)[31, 33, 48]. Such behavior is in sharp contrast with what would be expected in a system with a significant contribution of magnon-mediated spin currents. As typical magnonic group velocities are of the order of 10 nm/ps and smaller[49, 50], the resulting dynamics of the total spin current of conduction electrons and magnons would be heavily deformed and, for increasing $d_{\mathrm{AF}}$, exhibit an early electronic and delayed magnonic peak. The $d_{\mathrm{AF}}$-dependent relative delay would eventually leave our observation window (-0.4 … 0.8 ps). In addition, the recently observed ultrafast launching of magnonic currents, based on the spin Seebeck effect in metal-semimetal systems, would show a significantly slower dynamics, too.[51] As we do not observe any of these features and because our signals are not time-delayed with increasing $d_{\mathrm{AF}}$ we infer that the THz regime is dominated by a conduction-electron-mediated spin current. This conclusion is also consistent with a prior theoretical work[52] suggesting that the relevant characteristic spin-current decay length $\lambda$ in F|Pt systems is at THz frequencies determined by the mean free path of electrons, implying ballistic transport.

At GHz frequencies, two types of spin transport regimes – electronic (diffusive) and magnonic – may exist.[20, 21] Thickness-dependent spin-pumping experiments in F|FeMn($d_{\mathrm{FeMn}}$)|W[20] trilayers revealed non-monotonic S2C signals and, therefore, suggest a transition between spin transport regimes in FeMn. From our monotonic IrMn thickness-dependent S2C signals, we cannot disentangle electron and magnon contributions. If the magnonic component is not negligible, then a possible reason for why disentangling these contributions is more challenging for IrMn may be related with the shorter magnon characteristic lengths of 5 nm - as calculated in Ref. [53] - compared to 9 nm for FeMn. In that case, our data would infer that both the magnonic and electronic lengths are comparable (~2 nm).

Therefore, we can suggest the interpretation of the characteristic lengths $\lambda_{\mathrm{AF}}$, that differ by a factor of 4 between the THz and GHz data, as a consequence of different regimes of electronic spin transport: The ballistic regime at the THz frequencies where the electronic mean free path is the relevant quantity[31, 33], and the diffusive regime at the GHz frequencies characterized by the electron spin diffusion length[1] (typically longer than the mean free path[54, 55], implying slower dynamics). We cannot exclude though a magnonic contribution[14, 20, 21, 56] in the GHz experiments.

Finally, we focus on the reversely grown samples [Fig. 2 (d–f)]. If each stack NiFe|IrMn|Pt [Fig. 2 (d–f)] is a mirror image of its forward-grown partner Pt|IrMn|NiFe [Fig. 2 (a–c)], we would expect perfectly reversed signals since the spin and, thus, charge current flow is opposite and dominates over other THz-emission sources such as magnetic dipole radiation due to ultrafast demagnetization[33, 57, 58]. Because the excitation profile is nearly constant across the stacks[59], any deviations from this behavior indicate deviations from the ideal mirror image, which can in particular arise from the interface[32, 60] and its quality[40].

Although our simple model also well explains the reversely grown samples and they, therefore, provide values of $\lambda_{\mathrm{AF}}$ and $y_0$ very consistent with the forward-grown stacks [Fig. 2 (a–c)], we do observe a significant change of signal amplitudes for thin AF layers ($d_{\mathrm{AF}} < 2$ nm) quantified by $y_1 + y_0$ (Tab. S1). For instance, by comparing Fig. 2(b) and Fig. 2(e), the GHz data show a reduction of 2.5 (and smaller in other pairs), whereas the THz data differ by more than a factor of 9.

However, we find more irregularities present only in the THz regime. Unlike in the GHz regime, the THz data from reversely grown samples does not only differ by thickness-independent factors from their forward-grown counterparts, but it can also follow a non-monotonic trend, e.g. in the Pt-based trilayers [Fig. 2(a) vs

Fig. 2(d), or magnified in Fig. S4]. To test whether this might be an effect of growth-related differences in interfacial S2C, we make a linear combination of signals from the trilayer Pt|IrMn|NiFe (two interfaces present) and a control bilayer NiFe|IrMn (only F|AF interface present), shown in Fig. S4, which reasonably reproduces the non-monotonic trend from Fig. 2(d). This indicates that the signals from both interfaces changed their relative weights after reversing the growth without implying which one is more relevant, as detailed in the caption of Fig. S4.

Another striking signature of the interface impact that is manifested uniquely in the THz regime is observed at the thin limit of reversely grown Ta-based samples [Fig. 2(f)], in which we find no polarity switching with increasing $d_{\text{AF}}$. Such observation is unexpected considering the typical magnitude of the S2C conversion in Ta ($\theta_{\text{Ta}} \approx -7\%$ [41] comparing to small positive $\theta_{\text{AF}}$).

Interestingly, the dramatic reduction of the S2C amplitude, the non-monotonic $d_{\text{AF}}$-dependence or even the change of polarity of the S2C in the thin limit of the reversed-grown series, represented by $I_{\text{c,a}}$, is much more profound in the THz regime. It can be understood in terms of the spin memory loss (represented by a finite size layer with spin-dependent spin-flip scattering such as a finite spin diffusion length) and spin asymmetry (represented by an infinitesimally thin layer with spin-dependent electronic scattering, i.e. with spin-dependent mean free path) introduced by one of the IrMn interfaces, as argued in Refs. [44, 61, 62]. The intrinsic nature of the above two processes is very different and may impact the THz and GHz experiments differently, considering their distinct $\lambda_{\text{AF}}$. We note that the variations of $I_{\text{c,a}}$ due to different N materials are much smaller in the thick limit than in the thin limit. This may be an indication of the prevailing role of the IrMn/N interface.

In conclusion, we have shown that the ultrafast spin injection and conversion in IrMn is operative up to ~30 THz and currently limited by the pump pulse duration and detection bandwidth. The upper bound of the spin-to-charge conversion efficiency in IrMn, $(\lambda\theta^*)_{\text{IrMn}}$, amounts to roughly 10% of the conversion in Pt. The direct comparison of the THz to GHz regimes revealed that the characteristic length of the spin transport is 4 times larger at GHz frequencies. As the underlying mechanism, we suggest a dominating ballistic electron transport in the THz regime, compared to an electronic diffusive transport in the GHz regime mixed with an eventual magnonic contribution. We also showed that contributions of the interfaces to the spin-to-charge current conversion can be significant and even dominate the other conversion processes in the THz regime, thus making it useful in optimizing and engineering the ultrafast spintronic functionalities in antiferromagnets.

See the supplementary material for further details on the experiments.

## Acknowledgments

The authors acknowledge funding by the collaborative research center SFB TRR 227 "Ultrafast spin dynamics" (projects B02 and A05), the ERC H2020 through projects CoG TERAMAG/Grant No. 681917, the French national research agency (ANR) [Grant Number ANR-15-CE24-0015-01], the CEA's bottom-up exploratory program (Grant Number PE-18P31-ELSA), the European Research Council (ASPIN/Grant No. 766566) and the Czech Science Foundation (GA CR, Grant Number 21-28876J).

## Data availability

The data that support the findings of this study are available from the corresponding author upon reasonable request.

## Conflict of Interest

The authors have no conflicts to disclose.